\documentstyle[12pt,aaspp4]{article}
 
\newcommand{\sso}{$^{-1}$}

\begin{document}
\slugcomment{To be published in the Astronomical Journal, June 1999}
\title{Using Hubble Space Telescope Imaging of Nuclear Dust Morphology to Rule
Out Bars Fueling Seyfert Nuclei\altaffilmark{1}}
\author{Michael W. Regan\altaffilmark{2,3} 
\affil{Carnegie Institution of Washington,
Department of Terrestrial Magnetism,\\
5241 Broad Branch Road,
Washington, DC 20015}}
\author{John S. Mulchaey\altaffilmark{4}
\affil{The Observatories of the Carnegie Institution of Washington,
\\813 Santa Barbara Street,\\ 
Pasadena, CA 91101}}
\altaffiltext{1}{Based on observations with the NASA/ESA Hubble Space 
Telescope, obtained at the Space Telescope Science
Institute, which is operated by the Association of Universities for 
Research in Astronomy, Inc. under NASA contract No. NAS5-26555}
\altaffiltext{2}{Email-mregan@dtm.ciw.edu}
\altaffiltext{3}{Hubble Fellow}
\altaffiltext{4}{Email-mulchaey@ociw.edu}
\begin{abstract}
If AGN are powered by the accretion of matter onto massive
black holes, how does the gas in the host galaxy lose
the required angular momentum to approach the black hole?
Gas easily transfers angular momentum to stars in strong bars, making
them likely candidates.
Although
ground-based searches for bars in active galaxies using both
optical and near infrared surface brightness
have not found any excess of bars relative to quiescent galaxies,
the searches
have not been able to rule out small-scale nuclear
bars.
To look for these nuclear bars
we use HST WFPC2-NICMOS color maps to search
for the straight dust lane signature of strong bars.
Of the twelve Seyfert galaxies in our sample, only three have dust
lanes consistent with a strong nuclear bar.
Therefore, strong nuclear bars cannot be the primary
fueling mechanism for Seyfert nuclei.
We do find that a majority of the galaxies show an spiral
morphology in their dust lanes. 
These  spiral arms may be a possible fueling mechanism.

\end{abstract}
\keywords{galaxies: Seyfert --- galaxies: ISM --- galaxies: nuclei}
\section{Introduction}
One of the goals of AGN research is to try to
understand what transports the gas
from the large host galaxy scales
($\sim$1-10kpc) down
close enough to the nucleus ($<$1 pc) to fuel an AGN.
To accomplish this fueling 
the gas must decrease its angular momentum by 
several orders of magnitude.
How does the gas lose its angular momentum?
One possibility is that gas could lose angular momentum to stars in the 
shocks and gravitational torques 
caused by a bar-like gravitational potential 
(Sholsman, Frank, \& Begelman 1989).
If this is true, 
we might expect there to be more bars in active galaxies than in 
quiescent galaxies.
Surveys have looked for this excess of bars and none has been found
(Kotilainen et al. 1992; Zitelli et al. 1993; McLeod \& Reike 1995;
Alonso-Herrero, Ward, \& Kotilainen 1996; Mulchaey \& Regan 1997; 
Ho, Filippenko, \& Sargent 1997).
The lack of observational evidence for an excess of bars in
these survey
has not ruled out bars as a mechanism of fueling
active galaxies for several reasons.
One possibility is that there is an inherent difference between
active galaxies and normal galaxies in that not all normal galaxies
may contain a massive black hole.
In this case the lack of an excess of bars in Seyfert galaxies 
compared to
normal galaxies would not rule out bars as a fueling mechanism.
However, recent studies of quiescent galaxies using both ground-based
observations and the HST have revealed kinematic 
evidence for the presence of black holes
in almost every galaxy studied (Kormendy \& Richstone 1995).
Another possibility 
is that, due to a lack of resolution,
all studies to date have been unable to detect
small scale bars ($< \sim$3\arcsec),
which are all that is necessary to provide fuel to an AGN.

Almost all studies of AGN fueling in the past have focused on looking
for non-axisymmetric perturbations in the stellar surface density under the
assumption that these lead to perturbations in the gravitational potential.
However, this may not be the best way to search for small-scale nuclear bars.
Even in the near infrared 
the stellar surface density is subject to contamination by dust
extinction and regions of current star formation.
This contamination makes it hard to determine 
the gravitational potential from the stellar surface density.
Since the fuel for AGN is provided by the interstellar medium (ISM), another
avenue of investigation is to study the ISM directly.
Because the ISM in spiral galaxies is dynamically cold,
it responds strongly to perturbations in the gravitational potential.
If we can characterize the morphology of the ISM, it could provide
a more sensitive probe of the gravitational potential than the stellar
surface density morphology.

To search for the signatures of a nuclear bar 
using the ISM morphology
requires the resolution of 
the Hubble Space Telescope (HST). 
In this paper we present our observations of Seyfert
galaxies made with the 
Near Infrared Camera and Multi-object Spectrometer (NICMOS).
We combine the NICMOS observations 
with archival observations of the same galaxies made with
the Wide Field and Planetary Camera 2 (WFPC2) to form color maps.
We use these color maps to investigate the nuclear ISM morphology of
a set of Seyfert galaxies as a test of 
whether small scale bars fuel Seyfert nuclei.

\section{Using the ISM as a Probe of the Gravitational Potential}

\begin{figure}[htbp!]
\plotone{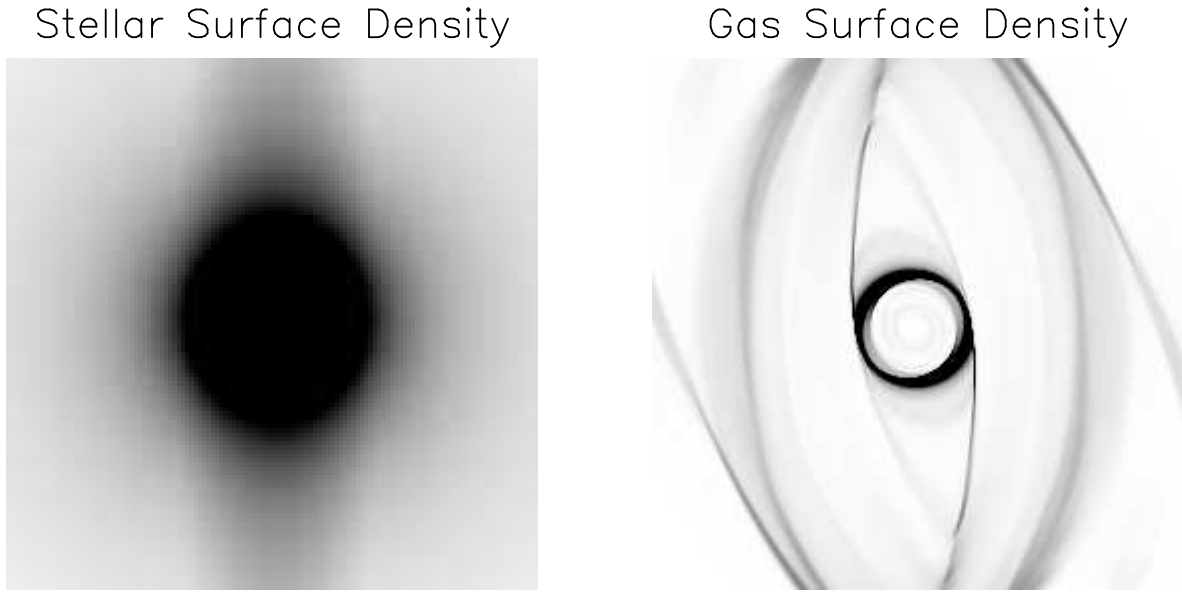}
\caption{a) The model stellar surface density of barred galaxy.
b) The model gas surface density of a barred galaxy. Notice how the
dust lanes along the leading edge of the bar are straight. This is the
signature of a strongly barred galaxy.}
\label{modelden}
\end{figure}

Since stars compose the vast majority of the mass in the central regions
of disk galaxies, it would make sense to search for perturbations in the
gravitational potential by looking at the stellar surface density.
The problem is that the stars in the old 
stellar disk have velocity dispersions of 
$\sim$40 km s\sso\ (Mihalas \& Binney 1981) which will tend to
wash out the effects of perturbations.
To complicate matters,
in the nuclear regions of galaxies the velocity dispersion of the
spheroid component is $>$100 km s\sso\
leading to even more smoothing of the surface density variations.
The low velocity dispersion of the ISM ($\sim$4 km s\sso\ in the local disk
of the Milky Way (Combes 1991)) will lead to much larger surface density
perturbations, making it a better probe of the gravitational potential.
To quantify the advantage of using the ISM surface density, we 
can compare the surface density variations of the two components
using a model galaxy.
For our model galaxy we determine
the stellar surface density by adding the contributions of 
three mass components: a disk, a bulge, and a bar.
The surface density of the Kuzmin-Toomre disk component, $\sigma_{d}$, is
\begin{equation}
\sigma_d(r)=(v^2_0 / 2 \pi G r)(1 + r^2/r^2_0)^{-1.5},
\end{equation}
where $v_0$ = 164 km s\sso\ and $r_0$ = 20 kpc.
The volume density of the bulge, $\rho_b$,  is
\begin{equation}
\rho_b=\rho_{cen}(1+r^2/r^2_{core})^{-1.5},
\end{equation}
where $\rho_{cen}$ is the central density of the bulge and $r_{core}$
is the core radius of the bulge.
We use a Ferrers ellipsoid for the bar density with a density,
$\rho_{bar}$,  of 
\begin{eqnarray}
\rho_{bar}=\rho_0(1-g^2) & \mathrm{for\: g} < 1\\
\mathrm{and}\nonumber\\
\rho_{bar}= 0 & \mathrm{for\: g} >=1,\nonumber
\end{eqnarray}
where $g^2 = y^2/a^2 +(x^2 + z^2)/b^2$ and $a$ and $b$ are the
semi-major and semi-minor axis, respectively.
To convert the volume densities into observable units we integrate
the volume densities of the bar and bulge
over $\pm$ 10 kpc in the
z direction to form surface densities.
We then sum the
three surface density components to give the total surface density
(Figure \ref{modelden}a).

We use the hydrodynamic models of gas flow in a barred potential from Piner, Stone
\& Teuben (1995) to find the gas surface density for this mass surface density.
The resulting gas surface density is shown in Figure \ref{modelden}b. 
The peak variation in the azimuthal gas surface density occurs at a 
radius of $\sim$2 kpc.
At this radius the azimuthal gas surface density variation is $\sim$20 to 1. 
At that same radius in the stellar surface density (Figure \ref{modelden}a)
the peak density variation with azimuth is $\sim$2.2 to 1.
Therefore, at a given level of sensitivity to azimuthal surface density
variations, the ISM surface density will reveal much
smaller non-axisymmetric perturbations than the
stellar surface density.

Detecting the ISM surface density variations at sub-arcsecond resolution 
cannot be done with conventional methods of observing the ISM in galaxies such
as millimeter interferometers and thermal infrared observations of dust.
Although the current generation of millimeter
arrays do have sub-arcsecond resolution,
they will not have the surface brightness sensitivity to detect
normal molecular clouds at this resolution until the arrival of the
Millimeter Array.
Thermal infrared observations by ISO of hot dust in Centaurs A did reveal a
the expected dust morphology of a strong bar (Mirabel et al. 1999) but 
the resolution of these observations was 3\arcsec which will only
resolve the dust morphology in relatively nearby galaxies.
An alternative method for observing the morphology of the ISM, is to look
for the dust component of the ISM in extinction.
One problem with using
dust extinction as a tracer of the ISM is that if the extinction is high
enough for the dust to become optically thick then the light will be dominated
by the foreground unreddened stars.
This leads to regions with large amounts of dust having relatively unreddened
colors.
This problem can be minimized by using a near infrared color as the long
wavelength color, which will provide a much larger dynamic range of dust
column depths that will be detectable in a extinction map.

Another advantage
of using the dust morphology to probe the potential instead of
the stellar surface density is that the problem of the
the stellar isophotes being influenced by the presence of dust
or regions of current star formation is avoided.
This effect can even be significant 
in the near infrared where the extinction due to dust
is much lower than in the optical.
A foreground dust feature with an optical extinction of A$_v$=2 leads to a 25\%
reduction in the transmitted light at 1.6\micron.
Stronger dust features are common in our color maps (see Figure \ref{cmaps}) 
which will lead to biases if
near infrared isophotes alone are used.
Biases will also be induced by regions of current star formation that
are not obvious in the NICMOS images alone.

Figure \ref{modelden}b shows that the dust 
morphology of a barred galaxy has
a very specific signature which can be used to infer the presence of a bar.
The dust lanes along the leading edge of the bar are always very
straight in strongly barred galaxies (Athanassoula 1992; Piner et al. 1995).
It is these strong bars that are proposed to drive
gas inward since in less strongly barred galaxies there may not be
a shock in the dust lanes (A92).
In galaxies with weaker bars, the dust lanes along the leading 
edge of the bar are curved, with the 
degree of curvature increasing as the bar weakens.
In general, it is hard to distinguish the dust morphology of a weak bar from 
the dust morphology of
high pitch angle spiral arms (Regan \& Vogel 1994).

These simple models suggest that the dust morphology
of a galaxy
can be a sensitive probe of its gravitational potential.
By searching for straight dust lanes we can determine if galaxies have
strong bars that are depriving the gas of angular momentum and thus
driving it inward.
Since we are directly probing the fuel of the AGN we are less prone to
biases caused by dust and star formation.

\section{Observations and Data Reduction}
\begin{figure}[htbp!]
\plottwo{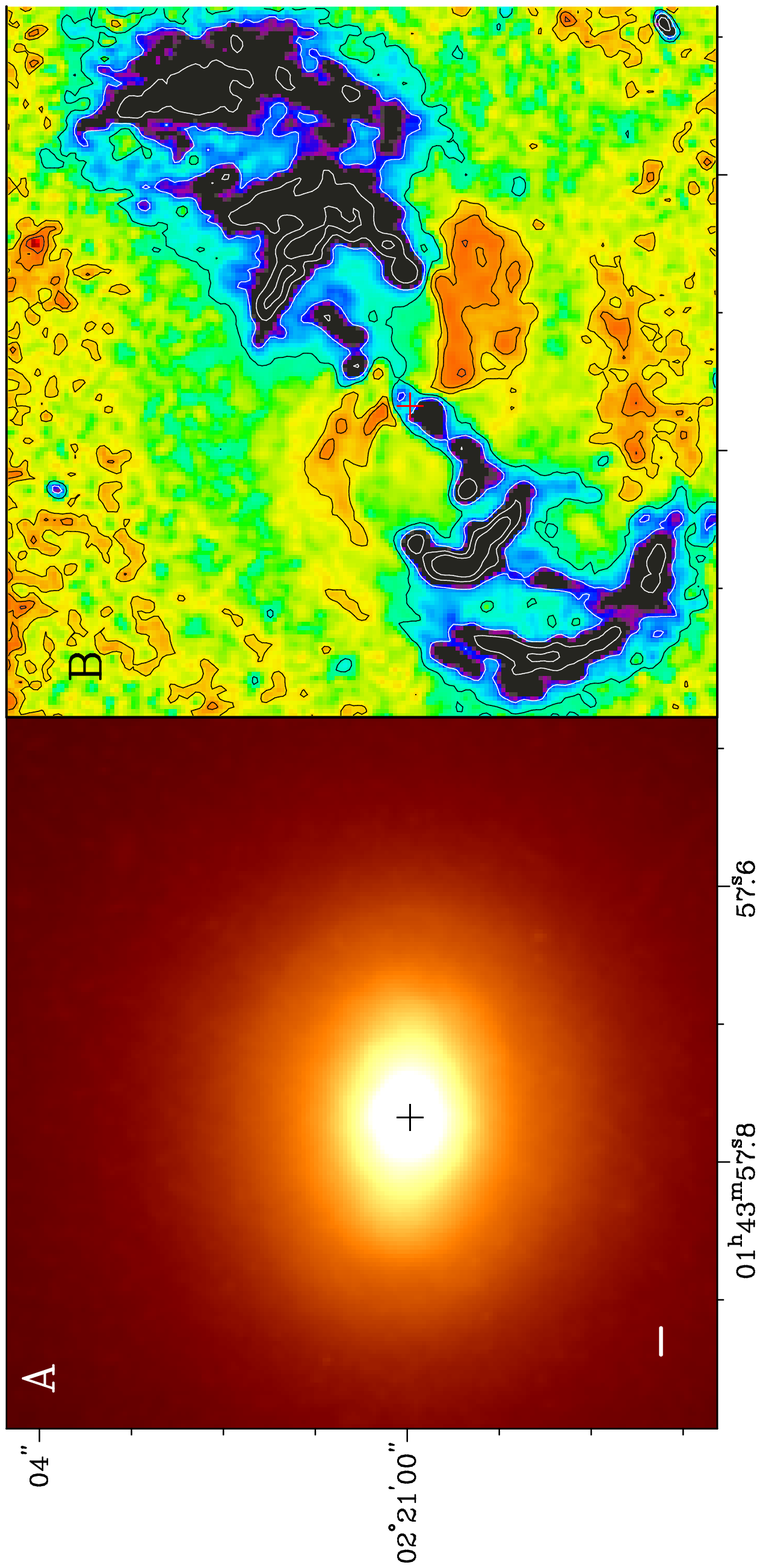}{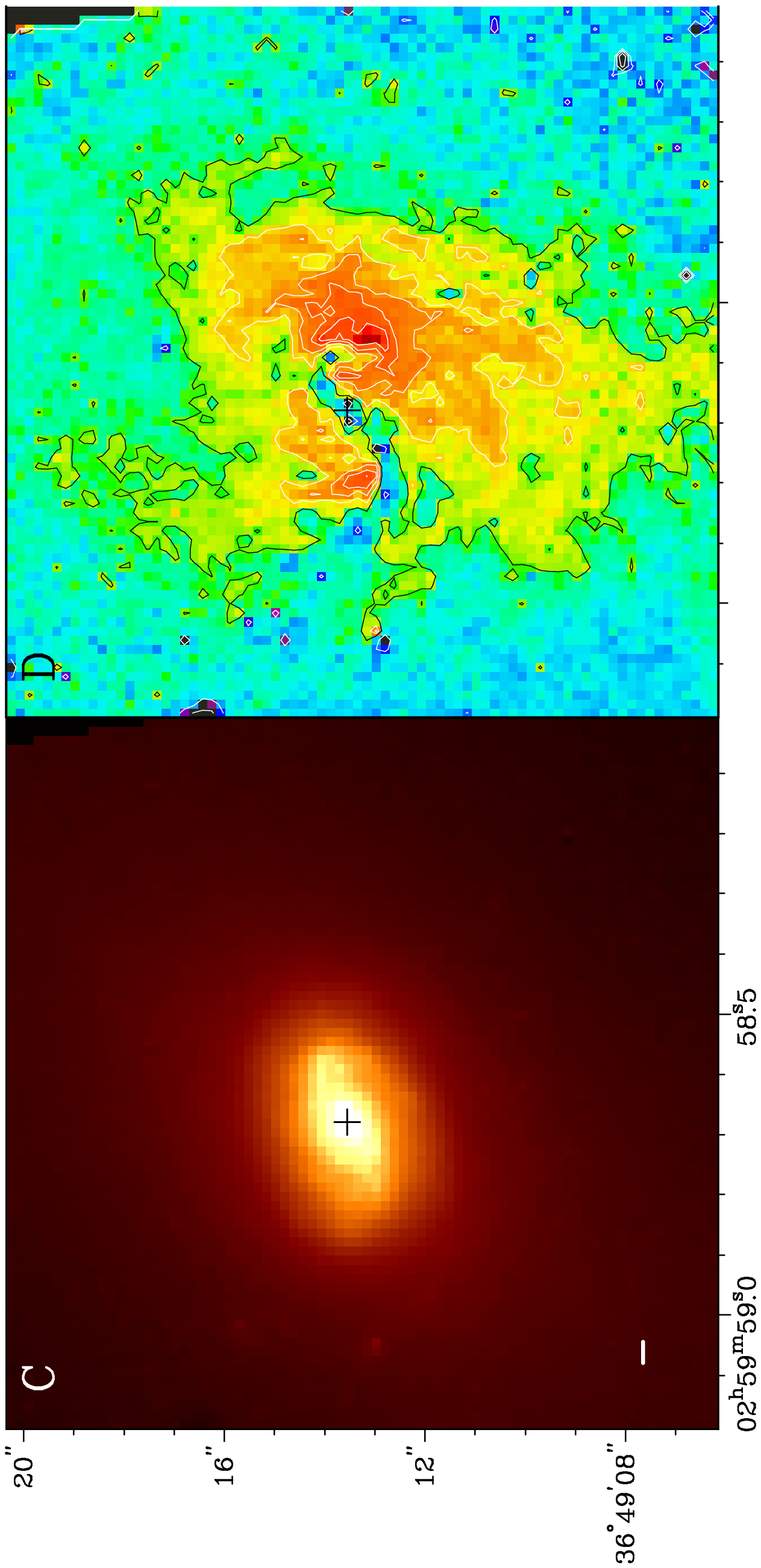}
\caption{Near infrared NICMOS 1.6\micron and color maps for the 
sample galaxies. Left column - NICMOS image taken at 1.6\micron using the
F160W filter; the cross marks the peak of the emission. The white bar 
in the lower left corner shows
a linear scale of 100pc assuming H$_0$=75 km s\sso\ Mpc\sso.
Right column - color map made by combining the NICMOS 1.6\micron\ image with
a WFPC2 F606W Image. The contours represent the red and blue color excesses
compared to the background galaxy. The contours of F606W$-$F160W blue 
color excess are at -0.2, -0.5, and -0.8. The contours of F606W$-$F160W red
color excess are at 0.3, 0.6, 0.9, 1.2, 1.5, and 1.9 magnitudes.ab) Mkn 573
cd) Mkn 1066 ef) NGC 1241 gh) NGC 1667 ij) NGC 3032 kl) NGC 3081 mn) 3516
op) NGC 3982 qr) NGC 5347 st) NGC 6300 uv) NGC 7582 wx) NGC 7743 }
\label{cmaps}
\end{figure}

Observations were made using the F160W filter on camera 2
of the Near Infrared Camera and 
Multi-Object Spectrometer (NICMOS) in the Hubble Space Telescope 
from the dates of 17 June 1997 to 18 March 1998.
The twelve galaxies in this sample are a subset of our larger
NICMOS sample of active and quiescent galaxies. 
These objects are the first
twelve in our sample observed by NICMOS for which HST observations
at optical wavelengths were available in the archive.
The F160W filter is an approximate match to ground-based near infrared 
H-band filters and has a central wavelength of 1.6\micron.
Our total integration time on source was either 704 or 640 seconds.
The shorter integration times were for those galaxies where we
dithered the telescope to allow for the removal of bad pixels.
The detector on Camera 2 is a 256$\times$256 HgCdTe detector with a
plate scale of 0.075 arc seconds pixel\sso\ yielding a field of view of
19\farcs2.
During the time period of the observations
the pipeline data reduction for NICMOS was still under
development leading to output of the pipeline process 
that was not the best possible
calibration of the data.
Therefore, we recalibrated the images using the latest flat fields, darks,
and non-linearity files.
The NICMOS images are shown in Figure \ref{cmaps}.

\clearpage
\plottwo{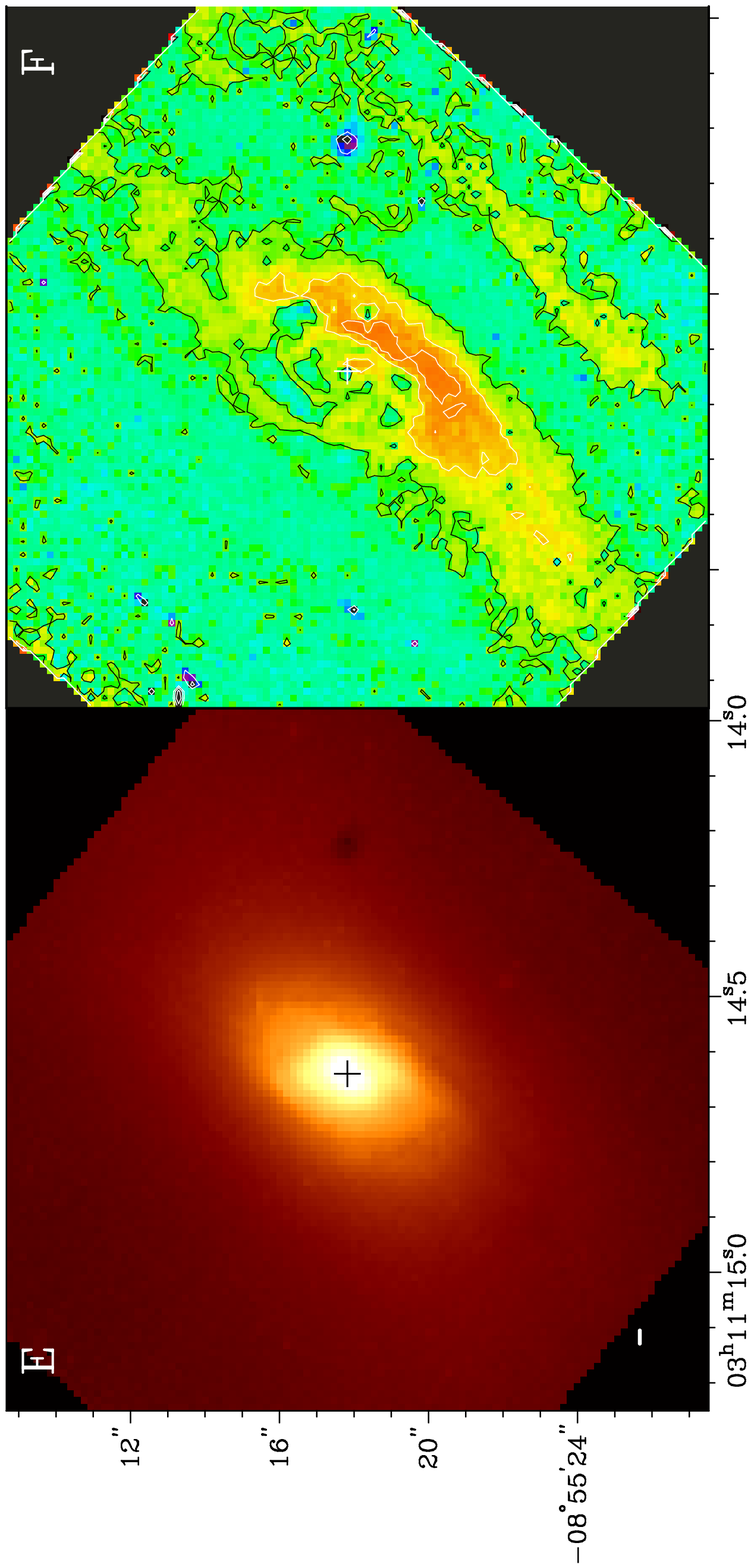}{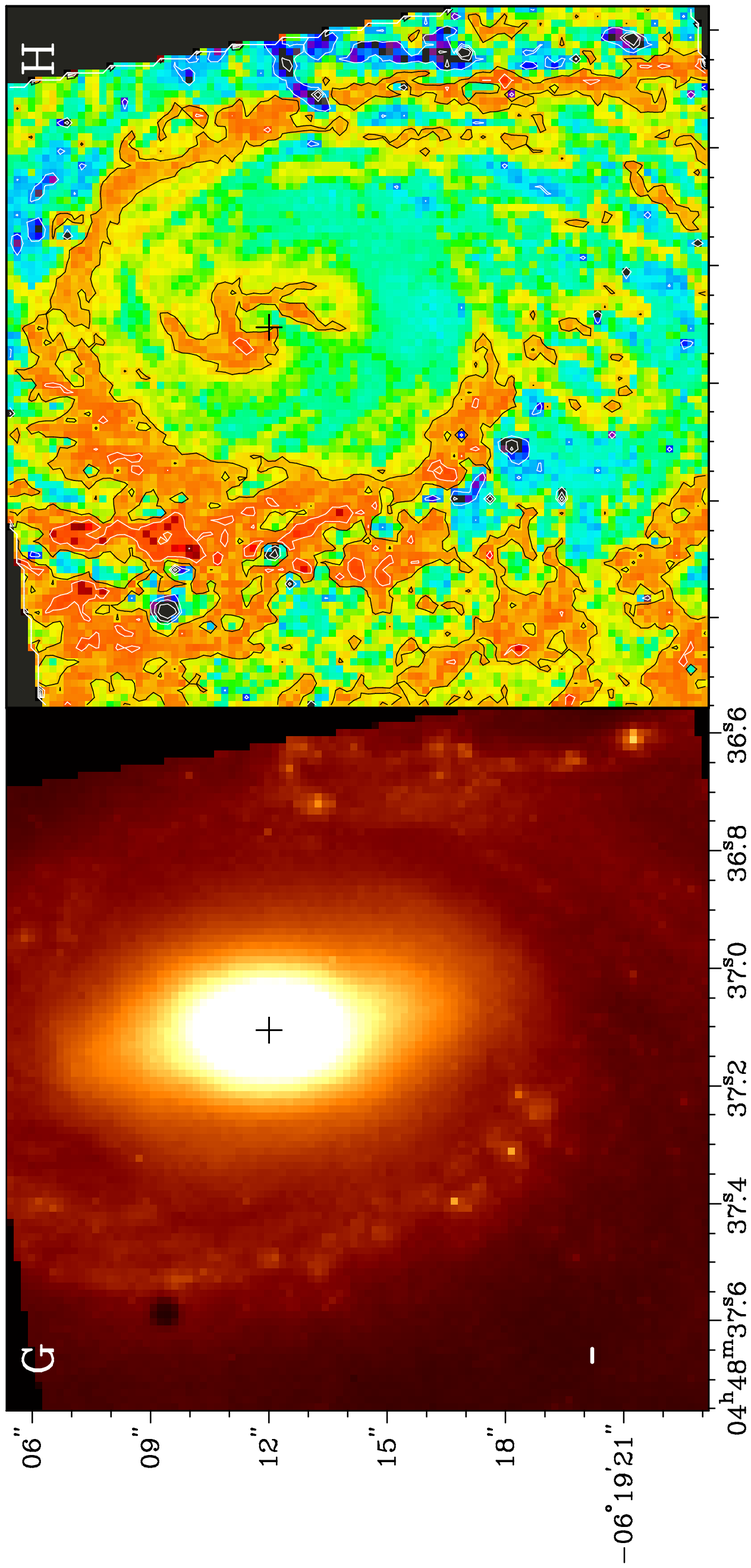}
\plottwo{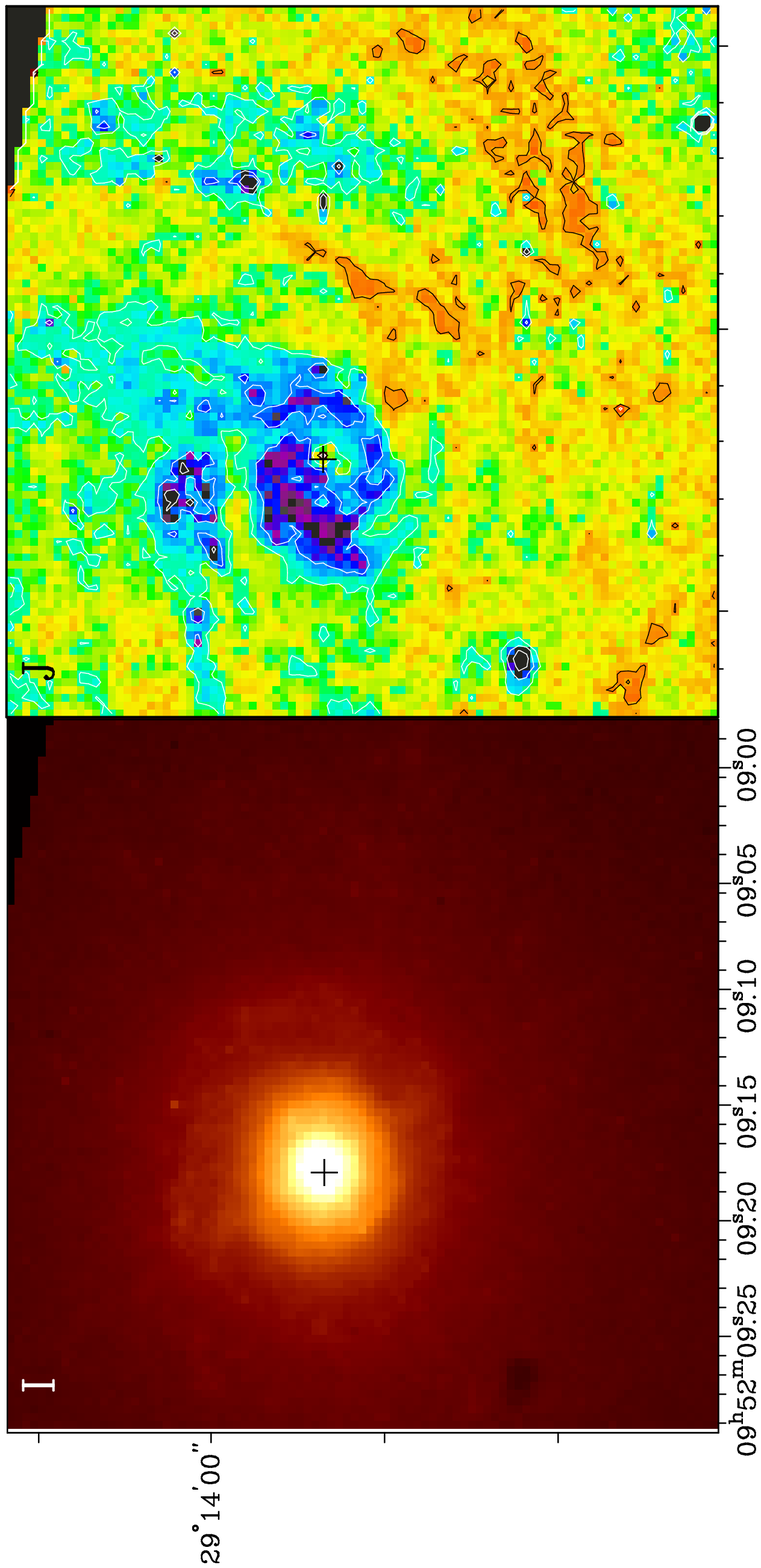}{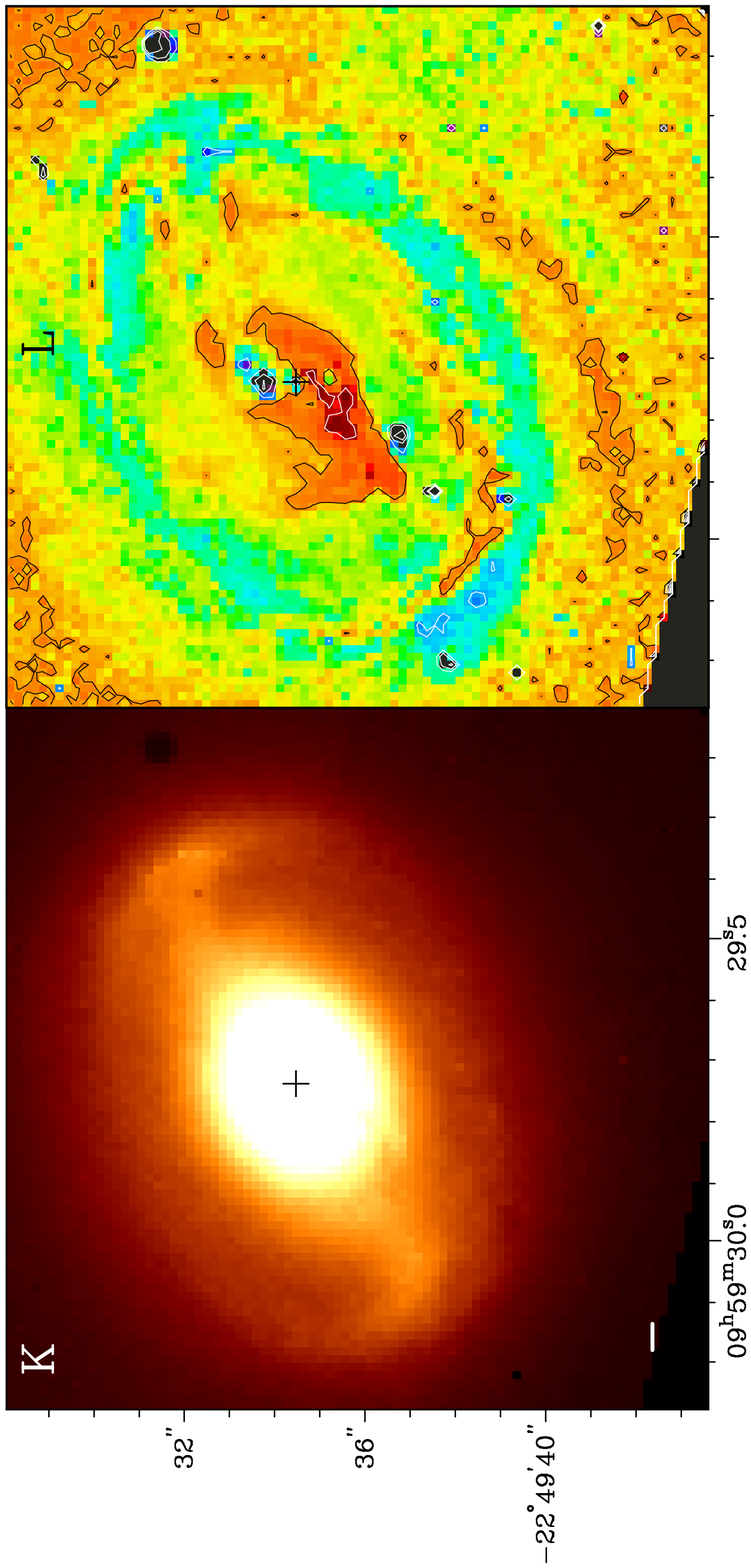}
\plottwo{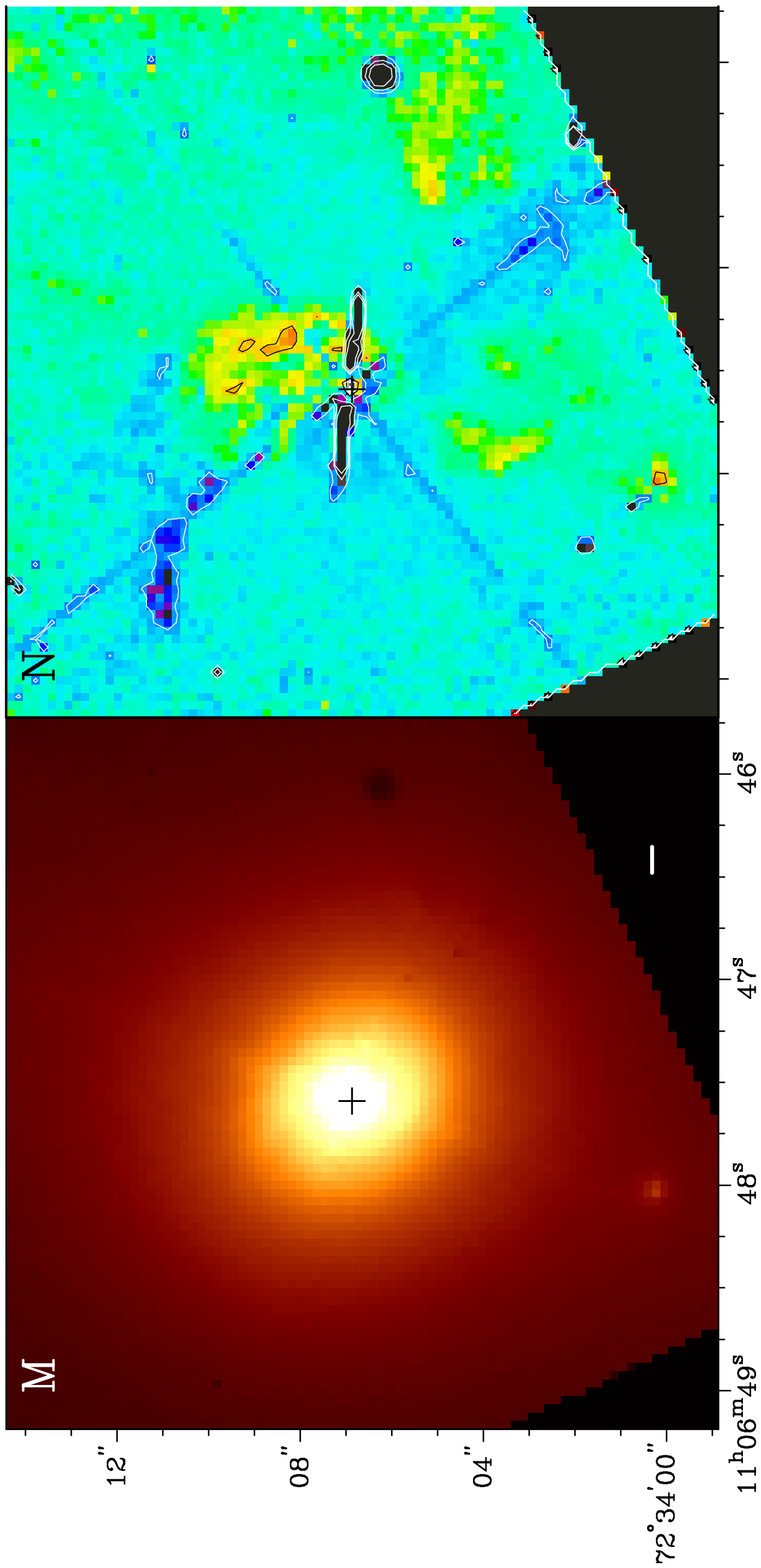}{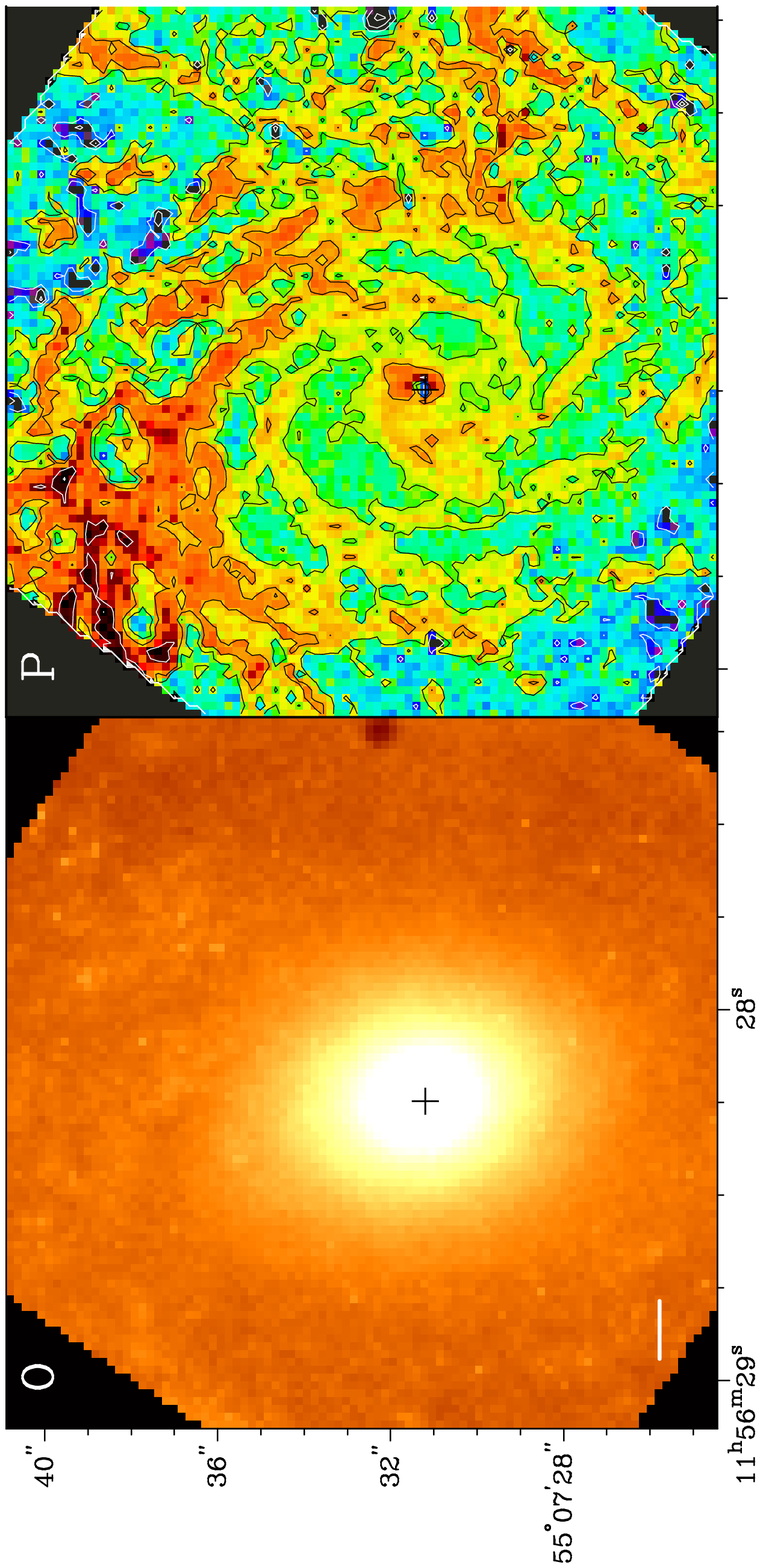}
\plottwo{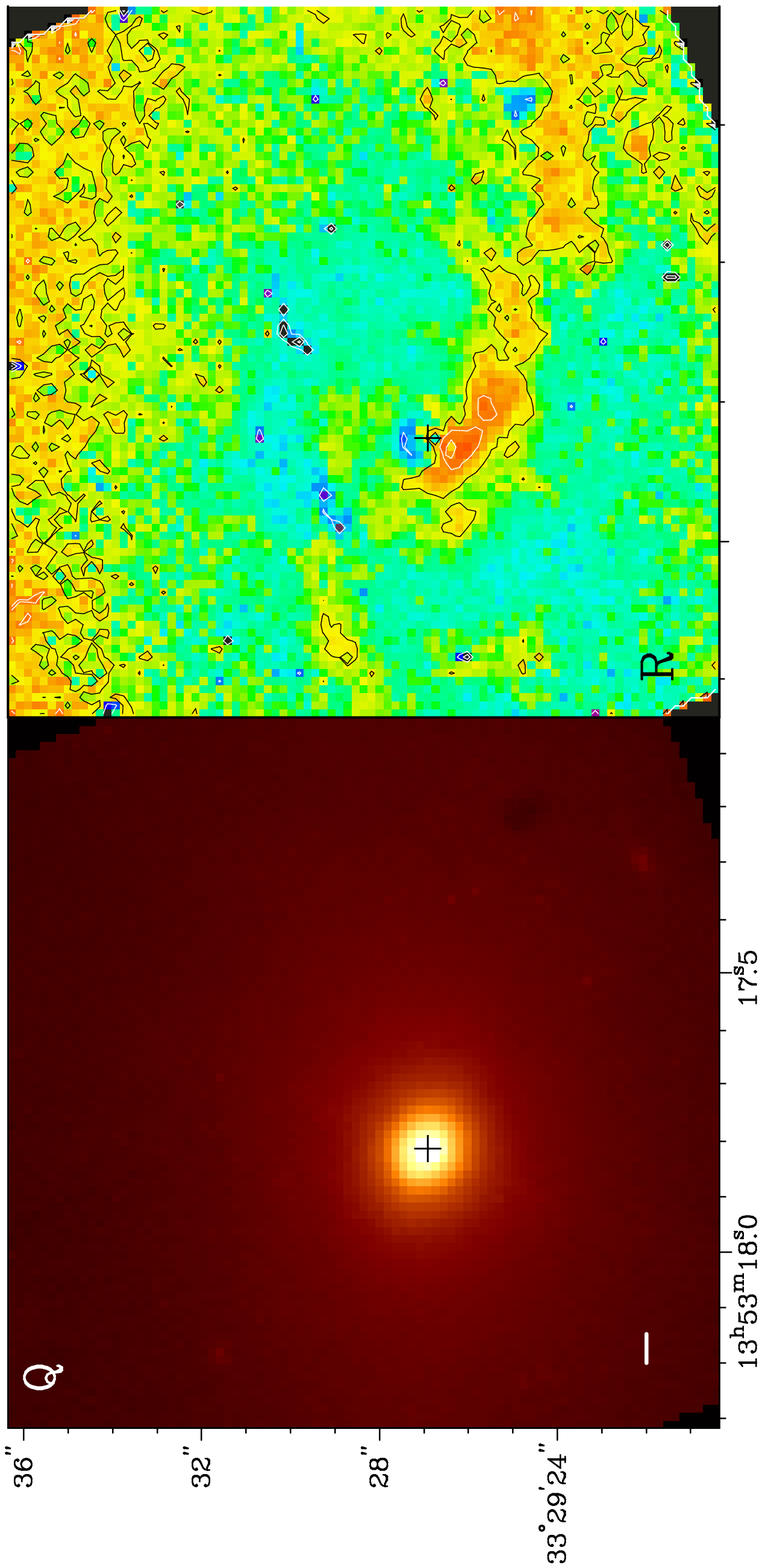}{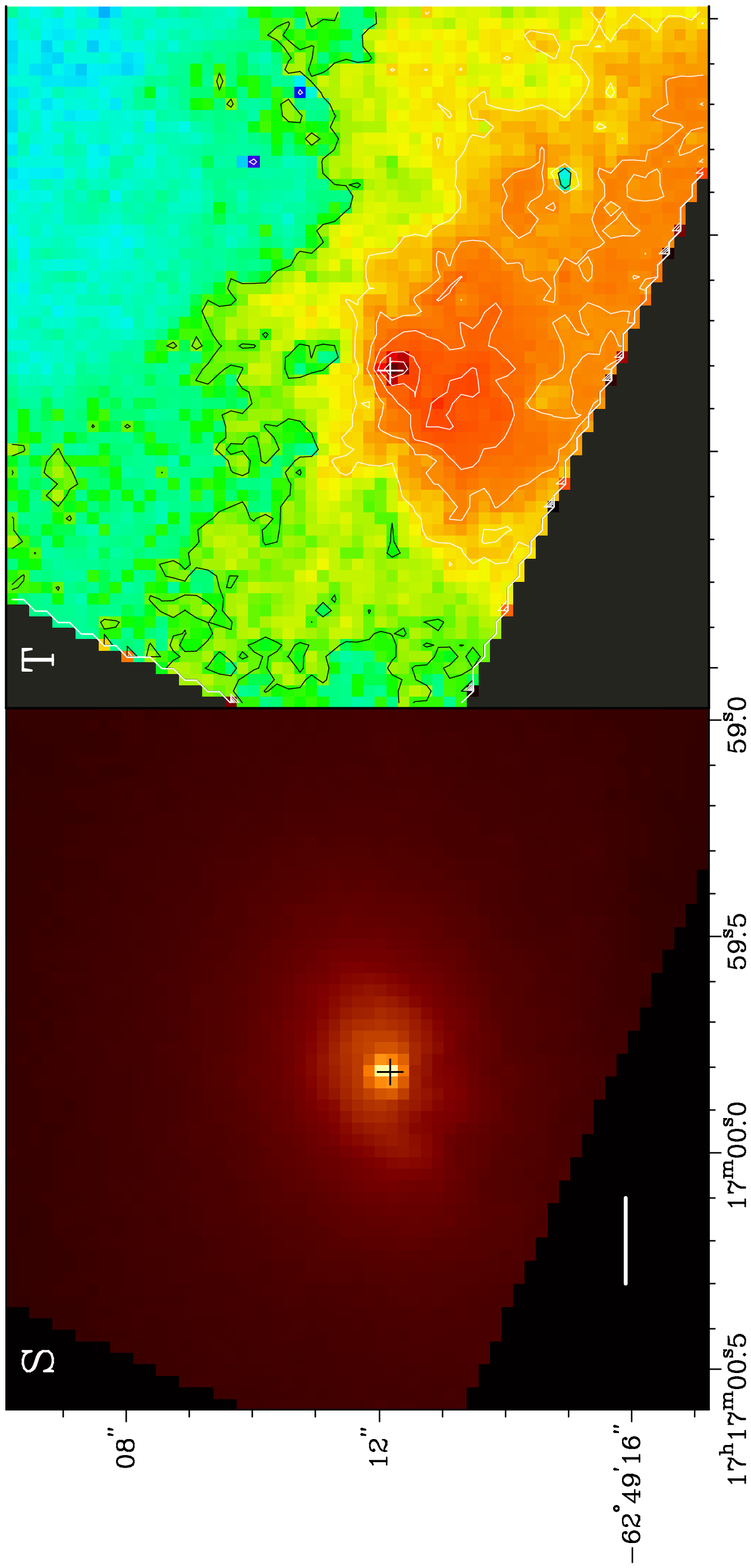}
\plottwo{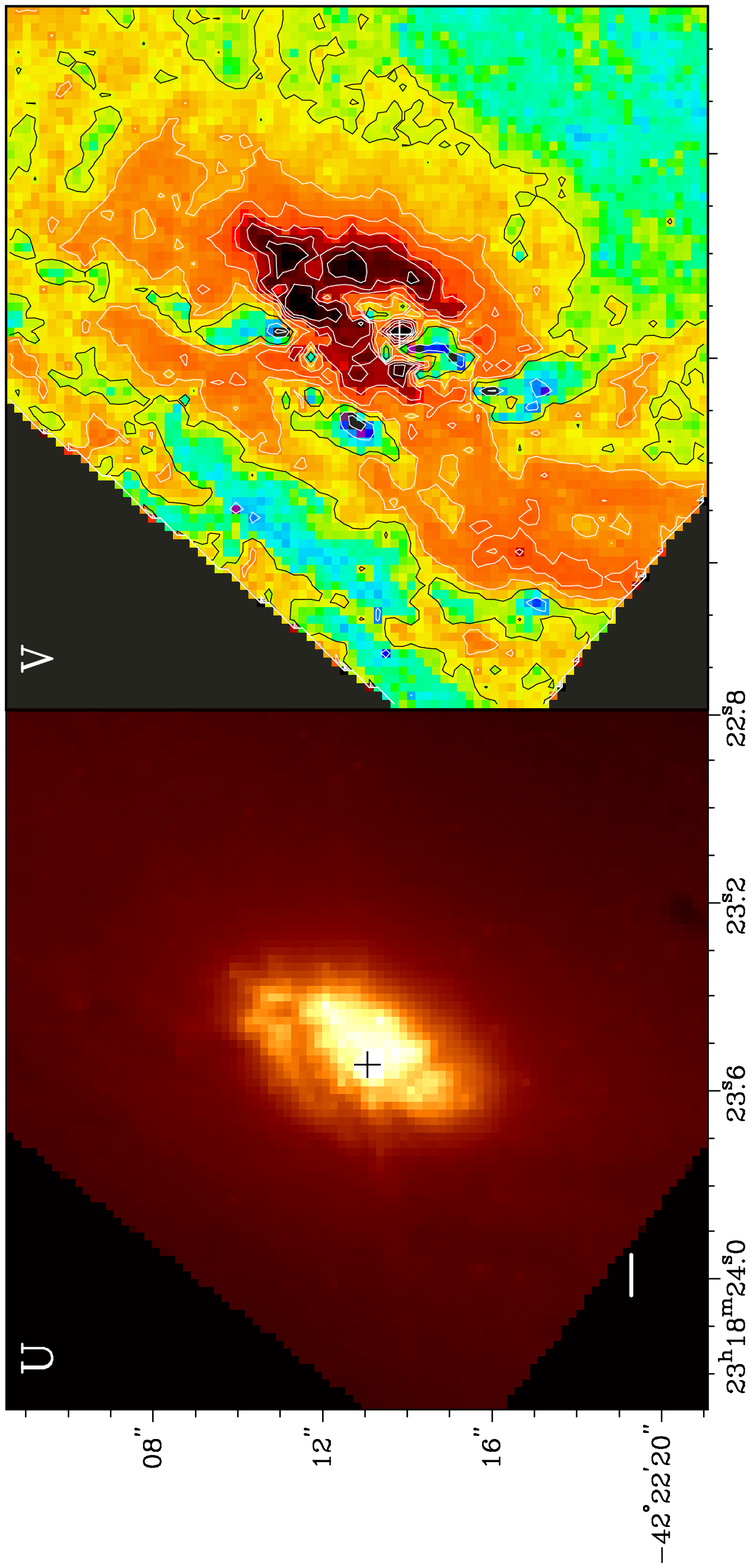}{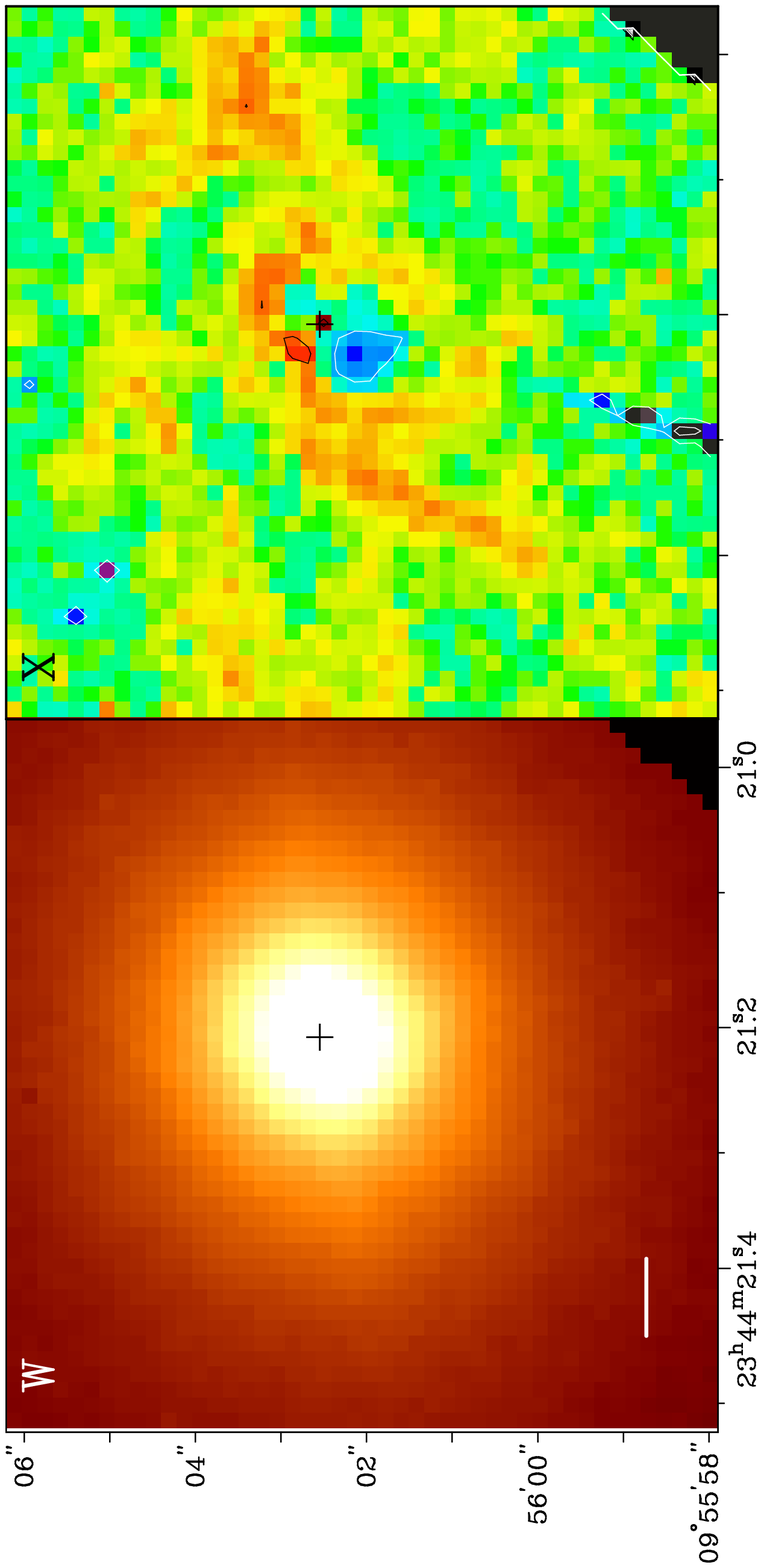} 

\flushleft
We obtained
optical images of the galaxies from the Hubble Data Archive.
These images were taken using the F606W filter using 500 second
exposures with the Wide Field and Planetary Camera 2
and have previously been published by Malkan, Gorjian, \& Tam (1998).
Since we were only interested in the regions that overlapped our NICMOS
images, we only used the planetary camera (PC) images.
We registered the NICMOS images to match the orientation and
plate scale of the PC images.
After registration there still was a 1-2\arcsec\ offset between the
NICMOS and PC images.
Therefore, 
we then performed a final shift to align the nucleus in the two images.
We then smoothed the PC images to same resolution as the NICMOS images and
divided the PC images by the NICMOS images to form color maps.
To make comparisons of extinction features in galaxies we need to know 
the color of the underlying stellar population.
For all the galaxies, except NGC 3982, we set the color excess to
be zero when the ratio of the count rate in the PC image to the count
rate in the NICMOS image was 0.09.
The underlying population in NGC 3982 is bluer than the other galaxies
leading us to set the color excess for this galaxy
to zero when the
ratio of the count rates was 0.11. 
The color maps for each of the sample galaxies are shown in Figure \ref{cmaps}.

\section{Results}

\subsection{Individual Galaxies}

\subsubsection{ Markarian 573}
The is a well studied Seyfert 2 galaxy with two ionization cones 
(Falcke, Wilson, \& Simpson 1998; Pogge \& De Robertis 1995).
The color map shows both blue color excess (light colors) 
and red color excess (dark colors) regions.
The region of the blue excess is delimited by two cones but there
is interesting structure within the blue regions.
Since the position angle of the cones in our color map 
(123\arcdeg) agrees well with the ground based ionization cone position
angle (124\arcdeg (Pogge \& De Robertis 1995)) the ionization cone is the blue
feature in our color map. The excess emission in the F606W filter
within the ionization cone is likely due to H$\alpha$ emission included in the
F606W filter bandpass.

What is unusual about this galaxy is that the red absorption features
connect to the blue emission features.
The strong red feature to the south of the nucleus can be seen to
directly connect to the very blue feature 2\arcsec\ west of the nucleus.
The weaker red dust lane to the north of the nucleus connects to
both a strong blue feature 0\farcs75 northwest of the nucleus and
faintly to the strong ridge of blue emission 2\arcsec\ east of the nucleus.
The two fainter outer red features can also be seen to connect to the
blue feature. 
The red feature 3\arcsec\ to the southeast of the nucleus connects to the
outer blue ridge with then shows a faint connection to a red feature to the
northeast of the nucleus.
The broad red feature just to the southeast of the nucleus shows a faint
connection to the broad blue feature east of the nucleus.
A possible reason for the connection between the features is that the blue
features are dust lanes illuminated from a central ionizing source.

Given the connection between the two types of features we can see that
Markarian 573 has four symmetrical spiral dust lanes.
Each dust lane traces about 180 degrees of arc as it spirals into
the central region.
The two outer dust lanes seem to have a higher pitch angle since they
reach the same radius as the two inner arms over the same angular extent
even though they start at about twice the radius of the inner arms.

\subsubsection{Markarian 1066}

The NICMOS image (Fig. \ref{cmaps}C) reveals what at first glance appears to be a nuclear bar
with a position angle of $\sim$135\arcdeg.
But our color map (Fig. \ref{cmaps}D) shows that the extension is caused by the regions of
current star formation just northwest and southeast of the nucleus.
The large amount of dust in this galaxy is quite evident as is the
overall spiral pattern to the dust.
A single broad dust lane seems to dominate the morphology.
This dust lane approaches the nucleus from the southeast and
reaches a maximum extinction about 1\arcsec\ west of the nucleus. 
A less well defined arm approaches the nucleus from the north and
reaches a maximum reddening about 1\arcsec\ east of the nucleus.
Just south of the maximum reddening the colors quickly change to blue.
It is clear from the figure that the dust is in a highly non-axisymmetric
distribution since the majority of the extinction is seen to the southwest of
the nucleus.

A string of blue features are seen east of the nucleus starting
about 1\farcs5 east of the nucleus and approach as close as 0\farcs5 to 
the nucleus.
They seem to connect to two blue features that are just east and west of
the maximum reddening of the western dust lane about 1\arcsec\ west of the
nucleus.

\subsubsection{NGC 1241}

The color map for NGC 1241 (Fig. \ref{cmaps}F) reveals an
overall dust morphology that is consistent with an inclined ring at the
terminus of bar dust lanes.
There is a red half ellipse close to the nucleus with a
minor axis of about 1\farcs5 and a major axis of $\sim$4\arcsec.
The other side of the ellipse does show in extinction but at a much lower
level.
This ellipse connects to
red features that lead off the image to the southeast and northwest.
The overall reddening to the
southwest of the nucleus is stronger
implying that this is the near side of the galaxy.

\subsubsection{NGC 1667}

The HST color map of NGC 1667 (Fig. \ref{cmaps}H)
 reveals a complex but relatively ordered dust morphology
with a nuclear spiral morphology.
The overall extinction is stronger to the northeast of the nucleus.
From the southwest a relatively straight dust lane approaches the nuclear
region until it intersects a nuclear ring.
Here it curves around until it joins the opposite dust lane approaching
the ring from the north.
The ring then continues around until it is south of the nucleus where it
then becomes hard to trace.
Inside the ring a multi-armed spiral pattern is seen near the nucleus.
These two arms are only roughly symmetrical and the center of symmetry
does not appear to be the nucleus. 
Blue, presumably star forming, regions are also visible in the map as 
clumps to the west of the western
dust lane and about 7\arcsec\ southeast of the nucleus near where the
eastern dust lane intersects the dust ring.

\subsubsection{NGC 3032}

Overall, NGC 3032 is much redder to the southwest and 
the nucleus is surrounded by a disk of blue emission in a spiral pattern.
In addition, a strong blue emission region is seen 3\arcsec\ northeast of
the nucleus.
Two red dust lanes can also be seen in the image.
The inner dust lane approaches from the north of the nucleus
and curves around the western side of the nucleus at a distance of
4\arcsec\ to connect to the blue disk southeast of the nucleus.
The other dust lane enters the image from the northwest and then exits the
image to the southeast. 
It may reenter the map northeast of the nucleus where a dust lane appears
to connect to blue features north of the nucleus.

\subsubsection{NGC 3081}

NGC 3081 was one of the best examples of a double-barred galaxy in the
survey of Mulchaey, Regan \& Kundu (1997).
The large bar had a radius of 37\arcsec\ while the nuclear bar had a radius
of 6\arcsec.
The nuclear bar can be seen in both the
NICMOS image (Figure \ref{cmaps}K) and the color map (Figure \ref{cmaps}L).
The most striking feature of the dust morphology is the ring of
extinction just outside of the nuclear bar.
Outside of this red ring feature in the color map 
is a more diffuse region of blue colors.
Both the red and blue features are stronger to the southwest of the 
nucleus.
A ring of star formation seen just outside of a bar is a common feature
and has been termed an inner ring (Buta 1986).
At the ends of the bar we can see faint regions of enhanced star formation
6\arcsec\ from the nucleus to the southeast and to the northwest. 

Within the nuclear 
bar along the leading edges of the bar there is a relatively wide
and dense region of extinction to the southeast.
On the other hand, the northwestern bar half shows much less extinction.
It does have a very blue, probably star forming, region 
just north of the nucleus.
The small separation of the star forming region from the active nucleus
($\sim$1\arcsec\ or 160 pc) has probably cased confusion in
ground based images and spectra of this galaxy.

\subsubsection{NGC 3516}
A single spiral dust pattern dominates the morphology of the S0 galaxy,
NGC 3516.
The main red dust lane emerges from the blue feature north of the
nucleus at a radius of $\sim$3\arcsec. 
It then appears to spiral around the nucleus and get closer until
it is lost in the diffraction effects $\sim$0\farcs5 from the
nucleus.
The extinction in the map is very asymmetric with the majority being
concentrated in north of the nucleus.
The saturated nucleus in the WFPC2 image 
creates several artifacts in the
color map.
These are the blue cross and the very blue east-west ridge.
A fainter red linear feature to the North and South is due to the
diffraction spikes in the NICMOS image.

There seems to be a connection between the dust lane to the north of the
nucleus seen in extinction and a blue feature to the northeast of the
nucleus. The morphology of the blue feature is not consistent with it arising
from star formation since it shows a filamentary structure.
The blue emission to the southwest of the nucleus also is more extended
than the diffraction spikes.
The diffuse blue emission $\sim$1\farcs5 southwest of the nucleus appears
to connect to the red excess seen 5\arcsec\ west southwest of the nucleus.

\subsubsection{NGC 3982}

The global extinction morphology of NGC 3982 is clearly a spiral pattern with
more extinction to the north of the nucleus (Fig. \ref{cmaps}P).
The spiral pattern appears to be multi-armed with there being
between two and four arms depending on how they are identified.
One spiral arm can be traced over 360\arcdeg\ degrees of arc and 
approaches to within the resolution of our images (0.15\arcsec\ or
10 pc).
Blue features are visible to the $\sim$8\arcsec\ 
southeast and northwest of the
nucleus on the outside of the dust lanes.
The dust lanes in the images also seem to be quite smooth as they 
near the nucleus compared to their more clumpy nature toward the
outer edge of the images (750 pc from the center).

\subsubsection{NGC 5347}

The main dust feature in the map is the broad 
slightly curved dust lane that
approaches the nucleus from the southwest and spirals around to
the east of the nucleus.
This is the type of dust lane that a strongly barred potential
will create.
There is also a blue region just to the north of the nucleus
at a distance of 1\arcsec\ (160 pc) which seems to connect to the
main dust lane.
A fainter extinction feature can also be seen to the northeast of
the nucleus that connects to north of the nucleus.

\subsubsection{NGC 6300}

The overall morphology of the extinction in NGC 6300 (Fig. \ref{cmaps}T)
shows no organization except that the
extinction is much stronger to the south of the nucleus than
to the north with the peak emission being just south of the
nucleus.

\subsubsection{NGC 7582}

The dust morphology of NGC 7582 (Fig. \ref{cmaps}V)
is quite complex in this galaxy but is dominated
by a dense ring of extinction.
This elliptical ring of extinction with a minor axis of $\sim$2\arcsec\
and a major axis of $\sim$3\arcsec\ surrounds the nucleus.
The ring is broken to the southeast by a chain of blue clumps.
The closest of these clumps is $\sim$0\farcs5 from the nucleus and the
chain extends out to 4\arcsec\ from the nucleus.
Two other blue clumps can be seen $\sim$3\arcsec\ east and $\sim$4\arcsec\
north of the nucleus.
A broad dust lane enters the image from the southeast and connects to the
extinction ring north of the chain of blue clumps.
As in Markarian 1066 the NICMOS image alone implies that there may be
a nuclear bar but the color map shows that this may be just a
combination of extinction and star formation perturbing the 
1.6\micron\ isophotes.

\subsubsection{NGC 7743}
The dust morphology of NGC 7743 (Fig \ref{cmaps}X)is consistent with a barred morphology.
The strongest dust lane is fairly straight starting to the southeast 
of the nucleus
but then it curves around the northern side of the nucleus.
It is then ends at clump of blue emission 0\farcs5 north of
the nucleus.
Two other fainter dust lanes can also be seen.
One south of the nucleus curves around to join the southeastern dust lane
 east of the nucleus.
Another dust lane seems to start north of the nucleus and curves around to 
the west and
south of the nucleus. 
A blue region of can be seen 1\arcsec\ southeast of the red nucleus.

\section{Discussion}

\subsection{Overall Dust Morphology}

We characterize
the nuclear dust morphology of the sample galaxies 
as barred, spiral, ring-like, or amorphous in Table \ref{dusttable}.
Dust morphologies consistent with a strong bar are found in only
three of the galaxies ruling out strong bar potentials as being the
primary fueling mechanism for all Seyfert nuclei. Even so,
these observations do not rule out bars as fueling mechanisms for the
more luminous active galaxies such as quasars and radio galaxies.
A weak bar cannot be ruled out in Mkn 1066 or NGC 3516 where the
dust morphology shows spiral arms with high pitch angles.
This type of dust morphology could
result either from spiral arms or a weak bar.

The other galaxies that exhibit spiral dust lanes
(Mkn 573, NGC 1667, NGC 3032, and NGC 3982) 
cannot have even a weakly barred nuclear potential because they either have
a multiple armed pattern or the pattern is tightly wound (low pitch angle).
A nuclear spiral pattern has been seen before in 
many galaxies (NGC 7252 - \markcite{W93}{Whitmore et al. 1993}; 
\markcite{M97}{Miller et al. 1997}, 
NGC 278 - \markcite{P96}{Phillips et al. 1996}, 
NGC 4414 - \markcite{TM97} {Thornley \& Mundy 1997},
NGC 5248 - \markcite{L98} {Laine et al. 1998} and in several Coma
cluster galaxies \markcite{CRD98}{Caldwell, Rose, \& Dendy 1997})
in both young blue stars and in dust extinction and thus appears to
be a common feature in disk galaxies.

In a recent paper on NGC 2207 Elmegreen et al. (1998) discuss acoustic noise
as a possible
formation mechanism for flocculent nuclear spirals disconnected from the
global spiral pattern.
This formation mechanism could explain the types of patterns we see
in NGC 1667 but it does not seem like it can explain the spiral patterns
in all the other galaxies.
Most notably in NGC 3982 we see that the nuclear spiral pattern is
continuous and connects to the larger scale spiral arms.

The two galaxies with a ring-like morphology (NGC 1241 and NGC 7582)
do show dust features that connect from the nucleus to the 
ring.
What makes them different than NGC 1667 or NGC 3081,
 which also show a ring of dust,
is that we cannot distinguish the morphology of the
dust interior to the dust rings.

The amorphous morphology of NGC 6300 in many ways is the most interesting.
It only takes a small non-axisymmetric perturbation of the potential to
induce a strong ordering to the gas morphology in a very short
period of time.
Even the weakest bar potentials 
show a very regular gas morphology (Anthanasoula 1992).
If the gas was to have enough mass for self-gravity to be important
one would expect that it would be unstable against multiple spiral
arm formation (Lin \& Shu 1964).
If not, then the acoustic noise mechanism might be expected to form 
a spiral pattern.
The lack of structure cannot be explained by a lack of linear 
resolution, since NGC 6300 is one of the closest galaxies in the sample.

\subsection{Dust Content}

\begin{figure}[htbp!]
\epsscale{0.8}
%\plotone{regan.fig3.ps}
\plotfiddle {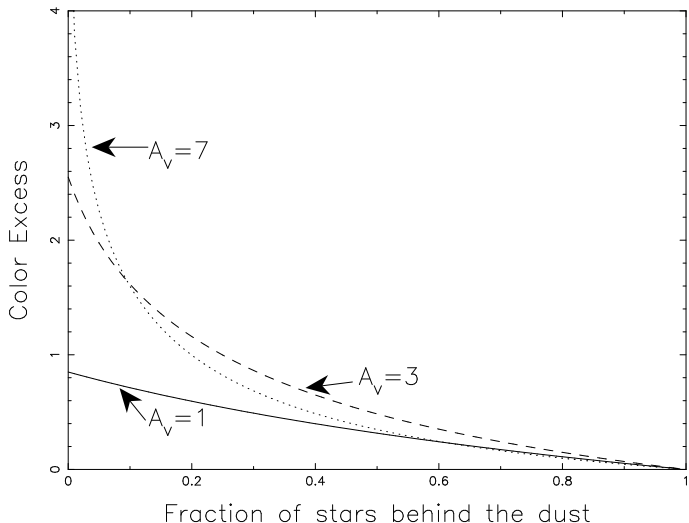}{2.2in}{-90}{35}{35}{-200}{200}
\caption{A plot of the color excess for a thin sheet of dust embedded in
plane of stars. The solid line, dashed line, and dotted line represent
an optical depth in V-band of 1, 3, and 5 magnitudes respectively. The color
excess is calculated using a V$-$H color which is a close match to the 
F606W$-$F160W color used in the paper. Notice how if the most of the dust
is in the mid-plane of the galaxy there is a maximum color excess of 0.6
magnitudes.}
\label{dustheight}
\end{figure}

Although, the reddening that we observe
for a given amount of dust is very
dependent on the location of the dust 
relative to the stars (Witt, 
Thronson, \& Capuano 1992), 
we can still
make some estimates of the total amount of dust and even the
geometry from the WFPC2-NICMOS color maps.
Because a given amount of dust produces the most reddening when the
dust is in front of all the stars,
we can derive lower limits for the dust
column depth using our single color.
The amount of reddening that can be
produced due to dust absorption is limited, if we make
the more realistic assumption that the dust and stars are mixed
together. 
The mixing of the dust and stars causes the stars on the far side of
the dust to contribute less light.
The reduction in light from the background stars causes
the observed colors to be dominated by the unreddened
foreground stars.
We can easily predict the
reddening that will be observed for a given ratio of foreground to
background stars, if we make the approximation that the dust scale height is small
relative to the scale height of the stars.
The reddening that would be observed for a various amounts of dust, 
based on the dust
characteristics of Bruzual, Magris, \& Calvet (1988) is
shown in
Figure \ref{dustheight}.
The observed reddening is reduced if the
dust back-scatters the light from the foreground stars 
so this Figure gives lower limits to the dust content.

Figure \ref{dustheight} shows that if the dust is in the mid-plane
of the galaxy, as we would expect near the nucleus where a large fraction of
the stars are in a extended spheroid or bulge, then the maximum color
excess we could observe would be $\sim$0.6 magnitudes.
The maximum extinction found in each of the galaxies is shown in
Table \ref{dusttable}. 
In most of the galaxies
a significant amount of dust must be above the mid-plane of the stars.
In fact, for three galaxies the dust must be in front of over 90\% of
the stars to account for the observed color excess.

The simplest explanation for the high color excesses is that we are 
observing a plane of dust inclined with respect to the bulge stars.
This would cause the
dust on the near side of the bulge to be in front of most of the
bulge stars.
This projection effect would only apply 
in regions that are neither too near
the nucleus nor too far from the nucleus.
Close to the nucleus the inclination of the disk does not result in 
enough change in the z height of the disk to affect the amount of
bulge stars seen in the background.
Far from the nucleus the contribution of bulge stars to the total light
is not high enough to significantly affect the colors.
If these very red colors are due to a projection effect then they should
be: only on one side of the galaxy, within the bulge region, and not too
close to the nucleus.
This projection effect may explain the color excesses in NGC 1241,
NGC 3081, NGC 3516, and NGC 5347.
It cannot explain the other galaxies and most 
importantly the three galaxies with
the highest color excesses (Mkn 1066, NGC 6300, and NGC 7582) do not
fit this view.
Mkn 1066 and NGC 7582 have very high color excesses on opposite sides of the
nucleus. 
A disk inside a bulge cannot produce high color excesses on different sides
of the galaxy.
In NGC 6300 the peak color excess is less than 0\farcs15 ($<$ 10 pc) 
from the nucleus.
This is too close for the projection effect to affect the color excess.

This leaves the possibility that the scale height of the dust 
is higher than the scale height of the stars.
In NGC 6300 not only must the scale height of the dust be very high but
the dust needs to be mostly in front of the galaxy.

\section{Conclusions}
Our color maps reveal a
 nuclear dust morphology that is
not consistent with strongly 
barred potentials being the only or even the
primary mechanism of driving gas into the nuclear region.
Only three of the 12 galaxies exhibit the 
straight dust lanes indicative of
a strongly barred potential.
Two of the galaxies have slightly curved dust lanes which could be
caused by a weak bar or oval distortion.
Our color maps allow us to recognize that what appear to be 
nuclear bars in
our NICMOS images of Markarian 1066 and NGC 7582 are 
more likely regions of nuclear
star formation.
The lack of nuclear bars in these Seyfert galaxies implies that there are
other mechanisms of fueling the nucleus or that Seyfert nuclei do not require
continuous fueling to remain active.

Another fueling mechanism could be the spiral dust lanes that
are quite common in our sample.
Many of the dust lanes can be traced to approach the nucleus
up to the limit of our resolution ($\sim$0\farcs15).
Mechanisms of both spiral arm formation and spiral arm fueling 
in the nuclear regions of galaxies need to be investigated.

We have shown how the dust morphology can be a sensitive
probe of the gravitational potential in a galaxy.
The lower velocity dispersion of the gas makes it respond much
more readily to perturbations.
In addition, using the dust morphology avoids
the problems of stellar isophotes being affected
by either dust extinction or current star formation.

Our color maps require that the dust in several of the galaxies
be distributed relatively high compared to the stars.
Only by putting the dust in front of most of the stars is it possible to
achieve the amount of reddening that we observe.
This high scale height dust is seen very near the nucleus of the galaxies
where the scale of the stars is large.

\acknowledgements
The authors would like to acknowledge helpful discussions with Johan Knapen,
Issac Shlosman, and Bruce Elmegreen and helpful comments from the referee.
Support for this work was provided by NASA through the Hubble Fellowship grant
\#HF-01100.01-98A awarded by the Space Telescope Science Institute, which is
operated by the Association of Universities for Research in Astronomy, Inc.
for NASA under contract NAS 5-26555.
Support for this work was also provided by NASA through grant number GO-07330
from the Space Telescope Science Institute, which  is
operated by the Association of Universities for Research in Astronomy, Inc.
for NASA under contract NAS 5-26555.

\clearpage
\begin{deluxetable}{lrrrrr}
\tablecaption{Galaxy Characteristics\label{galchar}}
\tablehead{\colhead{Galaxy}&\colhead{R.A.}&\colhead{Dec.}&\colhead{Hubble}&
\colhead{Systemic} & \colhead{Linear Scale} \\
\omit & \colhead{J2000} & \colhead{J2000} & \colhead{Type} & \colhead{Velocity\tablenotemark{a}} &\omit}
\startdata
Markarian 573 &01$^h$43$^m$57\fs8& 02\arcdeg 21\arcmin 00\farcs0 & SAB(rs)0+ Sy2 & 5174 km s\sso & 350 pc arcsec$^{-1}$\nl
Markarian 1066 &02$^h$59$^m$58\fs6& 36\arcdeg 49\arcmin 14\farcs0 &SB(s)0+ Sy2 & 3605 km s\sso & 240 pc arcsec$^{-1}$\nl
NGC 1241 &03$^h$11$^m$14\fs7&-08\arcdeg 55\arcmin 20\farcs0 &SB(rs)b Sy 2 & 4052 km s\sso & 270 pc arcsec$^{-1}$\nl
NGC 1667 &04$^h$48$^m$37\fs1&-06\arcdeg 19\arcmin 11\farcs9  & SAB(r)c Sy2 & 4547 km s\sso & 300 pc arcsec$^{-1}$\nl
NGC 3032 &09$^h$52$^m$08\fs2& 29\arcdeg 14\arcmin 10\farcs0  & SAB(r)0 & 1533 km s\sso & 100 pc arcsec$^{-1}$\nl
NGC 3081 &09$^h$59$^m$29\fs5&-22\arcdeg 49\arcmin 35\farcs0 & SAB(r)0/a Sy2 & 2385 km s\sso & 160 pc arcsec$^{-1}$\nl
NGC 3516 &11$^h$06$^m$47\fs5& 72\arcdeg 34\arcmin 06\farcs9  & SB(s)0 Sy1.5 & 2649 km s\sso & 180 pc arcsec$^{-1}$\nl
NGC 3982 &11$^h$56$^m$28\fs1& 55\arcdeg 07\arcmin 31\farcs0  & SAB(r)b Sy2 & 1109 km s\sso & 75  pc arcsec$^{-1}$\nl
NGC 5347 &13$^h$53$^m$17\fs8& 33\arcdeg 29\arcmin 27\farcs0  & SB(rs)ab Sy2&2335 km s\sso & 160 pc arcsec$^{-1}$\nl
NGC 6300 &17$^h$16$^m$59\fs2&-62\arcdeg 49\arcmin 11\farcs0  & SB(rs)b Sy2 & 1110 km s\sso & 75  pc arcsec$^{-1}$\nl
NGC 7582 &23$^h$18$^m$23\fs5&-42\arcdeg 22\arcmin 14\farcs1  &SB(s)ab Sy2 & 1575 km s\sso & 105 pc arcsec$^{-1}$\nl
NGC 7743 &23$^h$44$^m$21\fs5& 09\arcdeg 56\arcmin 04\farcs5  & SB(s)0+ Sy2 & 1710 km s\sso & 115 pc arcsec$^{-1}$\nl
\enddata
\tablenotetext{a}{All velocities in this table use the optical convention}
\end{deluxetable}

\begin{deluxetable}{lrr}
\tablecaption{Nuclear Dust Characteristics\label{dusttable}}
\tablehead{\colhead{Galaxy}&\colhead{Nuclear Dust Morphology}&{Maximum Color Excess}}
\startdata
Markarian 573 & spiral& 0.3 mag \nl
Markarian 1066 & spiral & 1.9 mag\nl
NGC 1241 & ring & 1.1 mag\nl
NGC 1667 & spiral & 0.9 mag\nl
NGC 3032 & spiral & 0.6 mag\nl
NGC 3081 & bar & 0.7 mag\nl
NGC 3516 & spiral & 0.5 mag\nl
NGC 3982 & spiral & 0.8 mag\nl
NGC 5347 & bar& 0.9 mag\nl
NGC 6300 & amorphous & 2.6 mag\nl
NGC 7582 & ring & 2.7 mag\nl
NGC 7743 & bar & 0.5 mag\nl
\enddata
\end{deluxetable}
\clearpage


\clearpage
\begin{references}
\reference {AH96} Alonso-Herrero, A., Ward, M. J., \& Kotilainen, J.
1996, \mnras, 278, 902
\reference {A92} Athanassoula. E. 1992, \mnras, 259, 345
\reference {B88} Bruzual, A. G., Magris, G., \& Calvet, N. 1998, \apj,
333, 673
\reference {CRD98}{Caldwell, N., Rose, J. A., \& Dendy, K. 1999, \aj, in press} 
\reference {e98} {Elmegreen, B. G., et al. 1998, \apjl, 503, 119L}
\reference {FWS98} Falcke, H., Wilson, A. S., Simpson, C. 1998, 
\apj, 502, 199
\reference {H97} Ho, L. C., Filippenko,A. V., \& Sargent, L. W. 1997, ApJ, 
487, 591
\reference {K95} Kormendy, J. \& Richstone, D. 1995, \araa, 33, 581
\reference {L63} Lin, C. C. \& Shu, F. H. 1964, \apj, 140, 646
\reference {PST95} Piner, B. G., Stone, J. M., \& Teuben, P. J. 1995, 
\apj, 449, 508
\reference {K92} Kotilainen, J. K., et al. 1992, \mnras, 256, 125
\reference {L98} Laine, S., Knapen, J. H., Perez-Ramirez, D., Doyon, R., \&
Nadeua, D. 1998, \mnras, in press
\reference {M95} McLeod, K. K. \& Reike, G. H. 1995, \apj, 454, 95
\reference {M98} Malkan, M. A., Gorjian, V., \& Tam, R. 1998, \apjs, 117,
25
\reference {M97} Miller, B. W., Whitmore, B. C., Schweizer, F.,
\& Fall, S. M. 1997, \aj, 114, 2381
\reference {M98} Mirabel et al. 1999, \aap, 341, 667
\reference {M97} Mulchaey, J. S. \& Regan, M. W. 1997, \apjl, 482, 135L
\reference {M97} Mulchaey, J. S., Regan, M. W., \& Kundu, A. 1997, \apjs, 110, 299
\reference {P96} Phillips, A. C., Illingworth, G. D., MacKenty, J. W.,
\& Franx, M. 1996, \aj, 111, 1566
\reference {PD95} Pogge, R. W. \& De Robertis, M. M. 1995, \apj, 451, 585
\reference {R95} Regan, M. W. \&  Vogel, S. N. 1995, \apjl, 21L
\reference {SFB89} Shlosman, I., Frank, J., \& Begelman, M. C. 1989, Nature,
338, 45
\reference {TM97} Thornley, M. D. \& Mundy, L. G. 1997, \apj, 490, 682
\reference {w93}{Whitmore, B. C., Schweizer, F., Leitherer, C., Borne, K., Robert, C. 1993,
\aj, 106, 1354}
\reference {w93b}{Witt, A. N., Thronson, H. A., Jr., \& Capuano, J. M., Jr. 1992 , \apj, 393, 611}
\reference {z93} Zitelli, V., Granato, G. L., Mandolesi, N., Wade, R.,
Danese, L. 1993, \apjs, 84, 185
\end{references}
\end{document}